# Bridging Structural Causal Inference and Machine Learning: The S-DIDML Estimator for Heterogeneous Treatment Effects

**Yile Yu*, Anzhi Xu**

**1.Zhejiang University of Technology, Hangzhou, 10337, China**

**2.Yunnan University of Finance and Economics,Kunming, 10689, China**

**Abstract:**In response to the increasing complexity of policy environments and the proliferation of high-dimensional data, this paper introduces the S-DIDML estimator—a framework grounded in structure and semiparametrically flexible for causal inference. By embedding Difference-in-Differences (DID) logic within a Double Machine Learning (DML) architecture, the S-DIDML approach combines the strengths of temporal identification, machine learning-based nuisance adjustment, and orthogonalized estimation. We begin by identifying critical limitations in existing methods, including the lack of structural interpretability in ML models, instability of classical DID under high-dimensional confounding, and the temporal rigidity of standard DML frameworks. Building on recent advances in staggered adoption designs and Neyman orthogonalization, S-DIDML offers a five-step estimation pipeline that enables robust estimation of heterogeneous treatment effects (HTEs) while maintaining interpretability and scalability. Demonstrative applications are discussed across labor economics, education, taxation, and environmental policy. The proposed framework contributes to the methodological frontier by offering a blueprint for policy-relevant, structurally interpretable, and statistically valid causal analysis in complex data settings.

## 1.Introduction

In the social sciences, the pursuit of causal understanding—rather than mere correlation—has long defined the discipline's scientific ambition. Whether assessing the effect of a trade policy on employment, the impact of educational reform on student performance, or the consequences of environmental regulations on firm productivity, researchers strive to isolate the effect of a particular treatment or intervention from the myriad of other confounding influences. At the heart of such endeavors lies causal inference, which seeks to answer the fundamental question: What would have happened in the absence of the treatment?

Economics, political science, education, and public policy have increasingly embraced quasi-experimental designs to approximate this elusive counterfactual.[1] Among them, the Difference-in-Differences (DID) framework has become a canonical tool.[2] It leverages temporal and group-based variation to identify causal effects under a critical "parallel trends" assumption.[3] The appeal of DID lies in its structural interpretability, transparency, and policy relevance, making it a mainstay in top journals and institutional evaluations alike.

However, the rise of high-dimensional data environments—where researchers can access hundreds of covariates from administrative records, digital platforms, or satellite imagery—has exposed the limitations of traditional causal tools.[4] Conventional DID approaches, often implemented via linear fixed effects models, may falter in the presence of nonlinear confounding, covariate imbalance, or heterogeneous treatment response.[5] In such settings, machine learning (ML) offers promising tools to flexibly model complex data structures, select relevant variables, and improve predictive accuracy.[6]

Yet, ML methods themselves are often prediction-oriented and lack the structural discipline needed for causal interpretation.[7] Bridging the divide between structure-driven identification and data-driven flexibility has become a central challenge for contemporary empirical research.[8] In response, a growing literature—most notably Double/Debiased Machine Learning (DDML)—has attempted to integrate the strengths of both paradigms.[9]

This paper situates itself within this growing effort to reconcile causal structure and machine learning innovation, focusing particularly on the evolution of DID and its modern extensions.[10] Through a comprehensive review and methodological synthesis, we propose a new estimator—S-DIDML—that structurally embeds DID within a residualized ML framework, enabling robust and interpretable causal inference in high-dimensional settings.[11]

## 2.Literature Mapping and Bibliometric Review

### 2.1The Evolution and Diversification of Difference-in-Differences in Social Science Research

Over the past three decades, the Difference-in-Differences (DID) approach has evolved from a relatively simple econometric tool to a cornerstone methodology for causal inference in applied social sciences.[12] First widely adopted in labor economics and public policy evaluation (Ashenfelter & Card, 1985), DID methods are now routinely employed to assess the causal impacts of reforms and shocks across domains such as education, environmental regulation, corporate governance, and innovation policy.[13] The conceptual strength of DID lies in its structural identification strategy, which exploits longitudinal variation between treatment and control units, assuming parallel trends in the absence of treatment.[14]

Our bibliometric analysis of 500 articles indexed in Web of Science (2000–2024) reveals an exponential growth in DID-related publications, especially after 2010. This growth reflects the method's diffusion beyond economics into interdisciplinary domains, including political science, environmental studies, and development research. The VOSviewer keyword co-occurrence map (Figure 1) illustrates how DID research has clustered around several thematic poles: green and environmental economics, anchored by terms such as pollution, green finance, and Porter hypothesis; corporate governance and financial policy, focused on ownership, incentives, and earnings management; innovation and productivity analysis, with keywords like urbanization, China, and total factor productivity; and public policy and human capital evaluation, integrating DID with propensity score methods, education, and healthcare outcomes.

The temporal cluster visualization from CiteSpace (Figure 2) suggests a clear trajectory of conceptual evolution.[15] Early DID studies (pre-2005) were tightly linked to labor market outcomes and macroeconomic shocks.[16] However, since 2015, the frontier has shifted toward applications in environmental regulation, technological innovation, and firm behavior under asymmetric information, often in the context of emerging markets such as China.[17] These newer applications frequently involve complex multi-treatment environments, time-varying exposures, and heterogeneous policy responses across sectors—conditions that test the limits of classical DID assumptions.

Methodologically, DID estimation has undergone substantial refinement.[18] Influential work by Abadie (2005) extended DID into the semiparametric domain, allowing for richer covariate structures via matching.[19] More recently, Callaway and Sant'Anna (2021) and Sun and Abraham (2021) introduced robust estimators for staggered adoption designs, correcting for treatment timing heterogeneity that plagues two-way fixed effects models.[20] These innovations address several longstanding concerns: violation of parallel trends across heterogeneous units; bias under dynamic treatment effects and anticipation; and negative weighting in TWFE regression models.

Despite these advances, traditional DID remains fundamentally limited in high-dimensional settings. Most DID implementations rely on linearity assumptions and low-dimensional covariate control, which become fragile when dealing with hundreds of potential confounders from administrative records, digital platforms, or geospatial data. Additionally, classical DID is typically estimated at the average treatment effect (ATE) level, with little consideration for treatment effect heterogeneity (HTE) across firms, regions, or demographic groups.

Recent attempts to augment DID with machine learning techniques—particularly in pre-processing stages using propensity score estimation, kernel matching, or covariate balancing—remain largely modular rather than integrative.[21] They seldom modify the core identification strategy or link to formal orthogonalization procedures, as found in Double Machine Learning (DML) frameworks.[22]

In summary, DID has achieved a remarkable degree of empirical relevance and institutional acceptance across disciplines.[23] However, the emerging landscape of complex, high-dimensional policy environments exposes critical limitations in classical DID frameworks.[24] To remain a credible tool for modern causal analysis, DID methods must evolve in two key directions: (1) integration with flexible, data-adaptive modeling approaches capable of handling nonlinearities and latent confounding, and (2) preservation of structural interpretability, ensuring that causal parameters remain transparent and policy-actionable.[25]

This methodological tension motivates the development of S-DIDML, a new estimator introduced in this study that embeds DID within a double residualized machine learning framework.[26] S-DIDML retains the core logic of temporal comparison while incorporating high-dimensional learning and orthogonalized estimation.[27] As we will show, this approach offers a principled and scalable path toward heterogeneity-aware, high-dimensional causal inference grounded in quasi-experimental logic.[28]

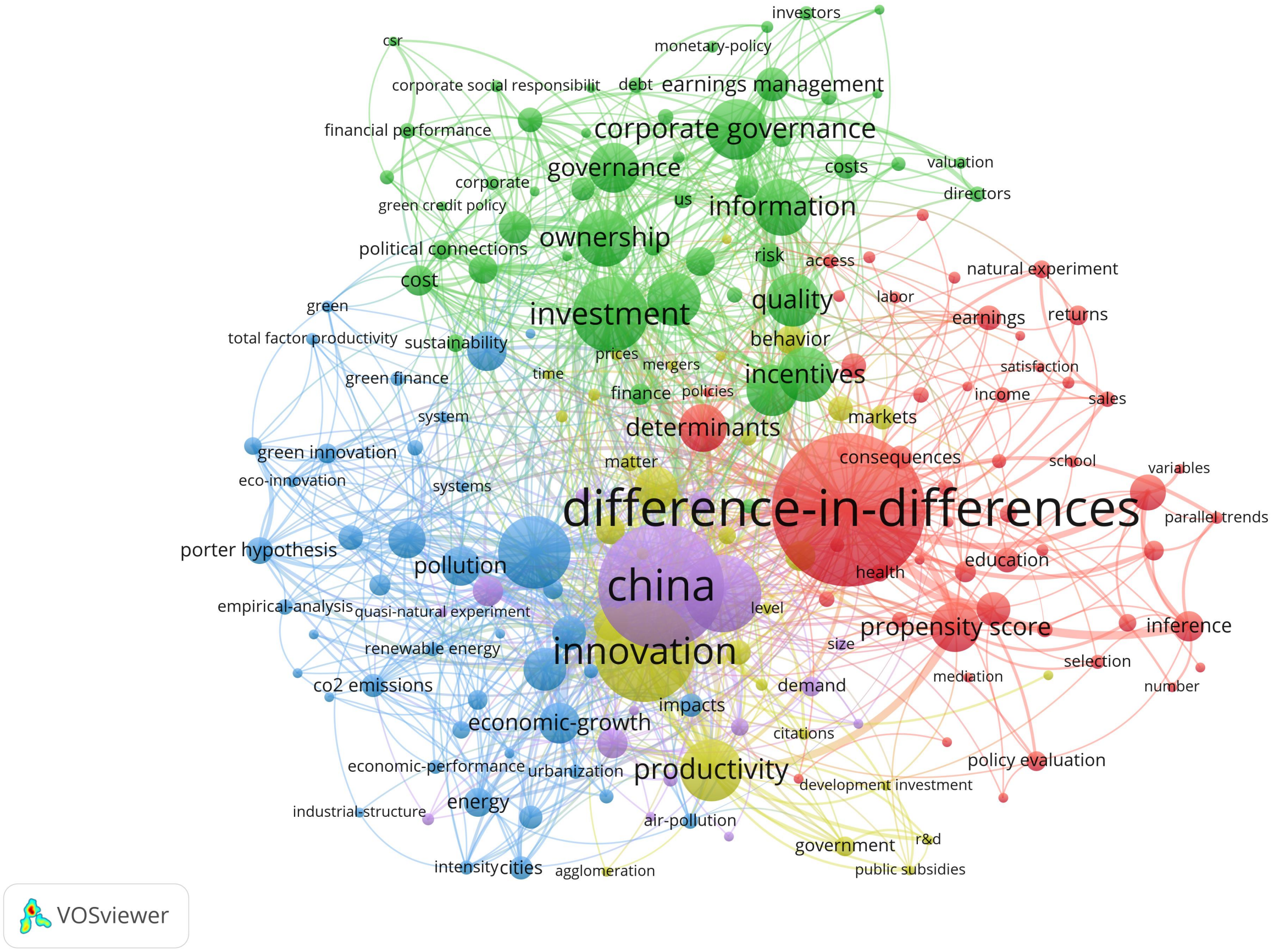

*Figure 1 Co-occurrence clustering map of keywords in DID-related research*

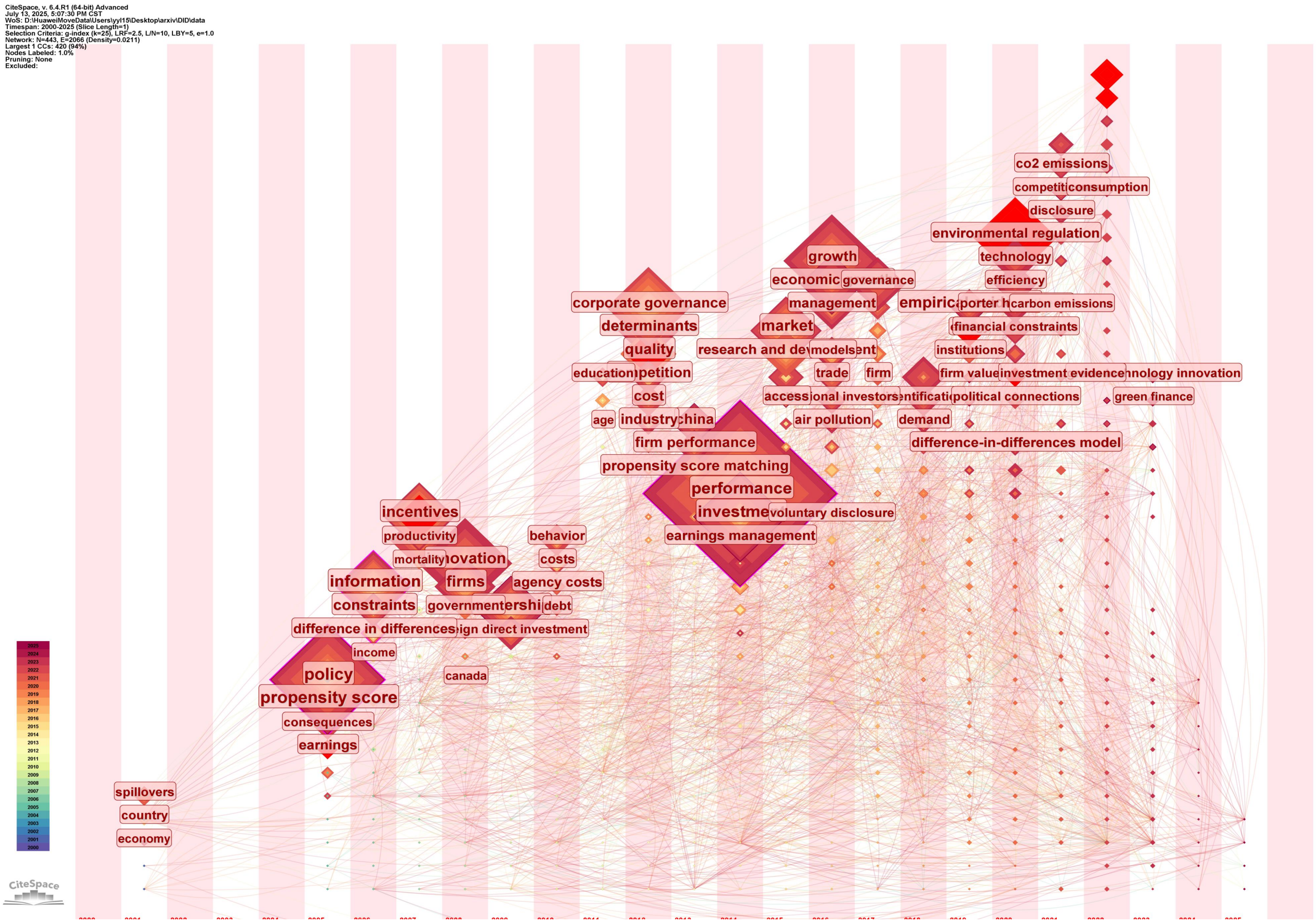

*Figure 2 Temporal evolution diagram of keywords in DID-related research*

## 2.2The Rise of Double/Debiased Machine Learning: From Orthogonalization to High-Dimensional Causal Inference

In response to the growing need for reliable causal inference in high-dimensional settings, the past decade has witnessed the rapid rise of Double Machine Learning (DML) and Debiased Machine Learning (DDML) frameworks.[29] Rooted in the econometric concept of orthogonalized moment equations and powered by modern machine learning-based nuisance parameter estimation, these methods have substantially expanded the empirical frontier for researchers dealing with complex treatment assignment, heterogeneity, and large-scale observational data.[30] DML and its variants provide a principled way to separate the prediction task (nuisance estimation) from the estimation of causal effects, by leveraging Neyman orthogonality, sample-splitting, and cross-fitting to obtain asymptotically valid estimators even in the presence of flexible, nonparametric models.[31]

The bibliometric analysis of 178 core papers indexed in Web of Science (2016–2024) confirms the explosive growth of this literature.[32] The VOSviewer co-occurrence network (Figure 3) centers around key themes including causal inference, Neyman orthogonality, cross-fitting, efficiency, and heterogeneous treatment effects, all closely linked with practical implementation strategies such as lasso, random forest, double robustness, and partially linear models.[33] Meanwhile, the CiteSpace temporal map (Figure 4) highlights a condensed knowledge burst in 2019–2022, during which seminal works formalized inference after machine learning (Chernozhukov et al.[34], 2018), introduced double/debiased orthogonal estimation (Newey & Robins, 2018), and developed scalable estimators for average and conditional treatment effects (Athey & Wager, 2019; Farrell et al., 2021). These methods have now been codified into a general paradigm: use ML to estimate nuisance components, and then debias the effect estimator via orthogonalization.

At the theoretical level, DDML builds on semiparametric efficiency theory, combining flexible first-stage ML tools

(e.[35]g., penalized regression, tree-based ensembles, deep nets) with second-stage doubly robust score functions. Estimators are designed to satisfy conditions such as local robustness, asymptotic linearity, and root-n convergence, making them particularly suited for policy evaluation under approximate sparsity or nonlinear selection. A critical innovation is the use of cross-fitting to mitigate overfitting bias in high-capacity learners, enabling the use of complex models like neural networks or gradient boosting within a valid inferential framework.

In empirical applications, DDML has rapidly diffused into fields such as health economics, education policy, taxation, and labor market discrimination, often in the context of heterogeneous treatment effects or partial identification.[36] For instance, Athey, Tibshirani, and Wager (2019) propose Causal Forests to estimate treatment effects conditional on covariates, while Chernozhukov et al.[37] (2020) extend DDML to quantile regression and instrumental variables. These contributions have redefined the scope of credible causal inference, enabling researchers to shift from estimating a single average treatment effect toward recovering rich heterogeneity structures and distributional effects, all while preserving valid statistical inference.

Nevertheless, current DDML applications still face several constraints.[38] First, many frameworks are cross-sectional or static, lacking explicit integration with panel designs, staggered treatments, or temporal structures typical of Difference-in-Differences research.[39] Second, the adoption of ML within causal inference remains primarily focused on prediction-quality improvements rather than structural causal modeling.[40] There is limited attention to embedding domain-specific identification strategies (e.g., timing, policy eligibility rules, group heterogeneity) within the estimation process. Finally, while DML allows for flexible controls, it does not by itself address the interpretability challenge: the black-box nature of some ML learners can obscure the causal estimand and undermine transparency in applied policy contexts.

These challenges underscore the necessity of an integrative approach—one that unifies the structure-driven identification logic of quasi-experimental designs with the data-adaptive capacity of machine learning. The S-DIDML framework proposed in this paper is precisely such an effort. By embedding the Difference-in-Differences design into a DDML architecture, we enable researchers to retain structural interpretability while achieving statistical robustness in high-dimensional, dynamic, and heterogeneous settings.

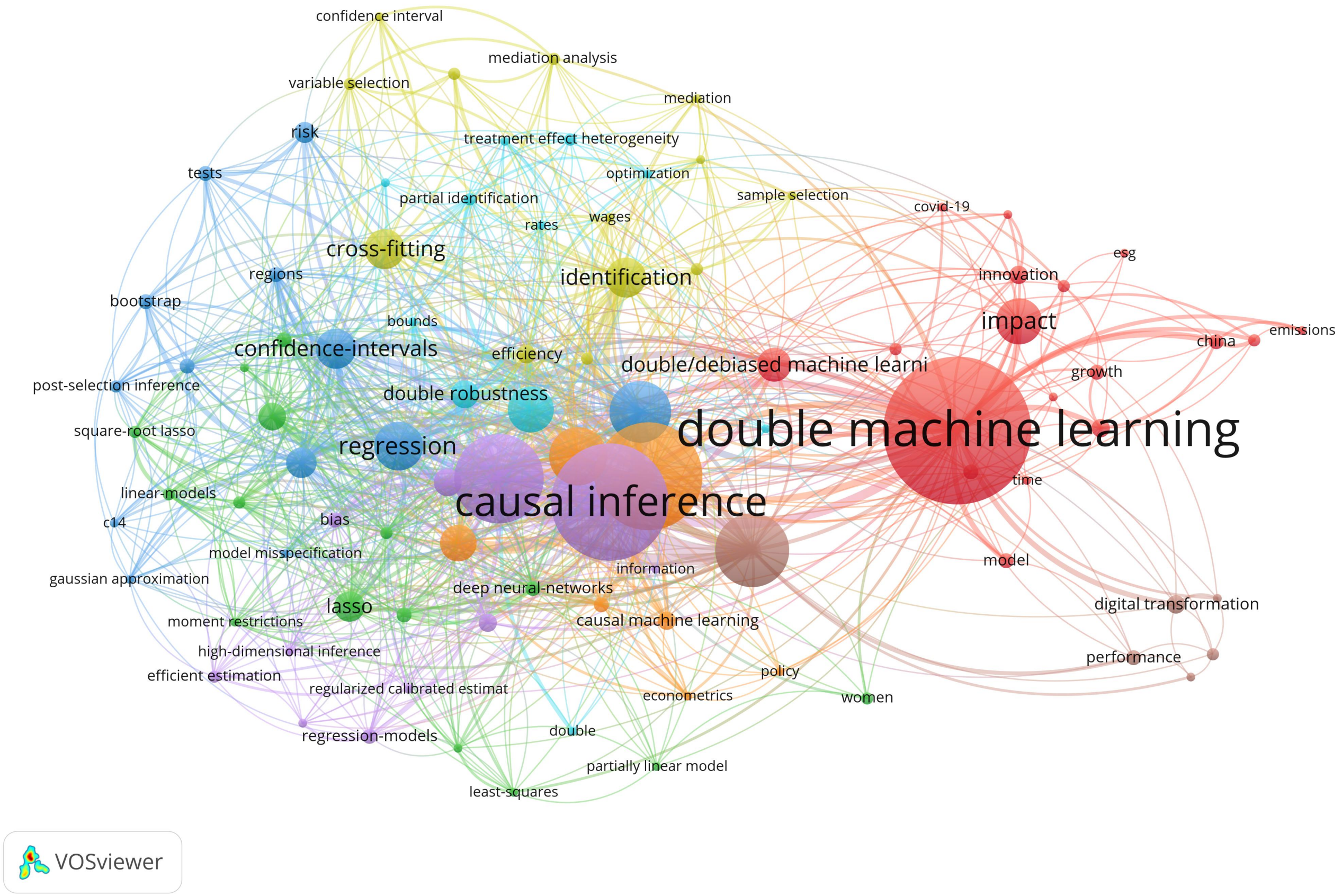

*Figure 3 Keyword co-occurrence clustering map of DDML-related research*

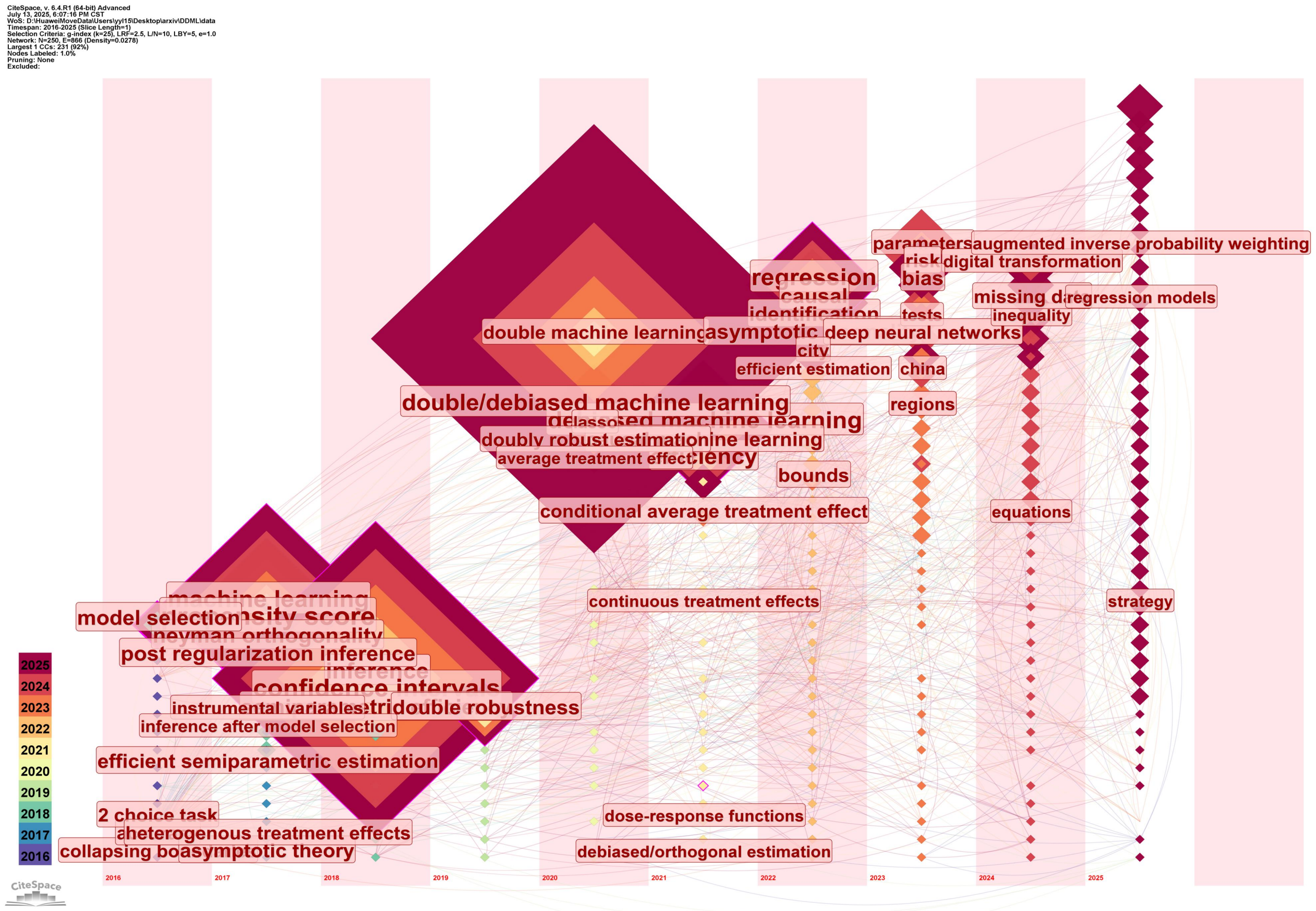

*Figure 4 Temporal evolution diagram of keywords related to DDML research*

**2.3Structural Causal Inference Meets Machine Learning: Toward a Unified Framework for Transparent and Scalable Policy Evaluation**

The increasing availability of high-dimensional observational data has amplified the need for causal inference methods that are both structurally interpretable and computationally scalable. This has spurred a growing body of research at the intersection of structural causal inference and machine learning (ML), which seeks to preserve the clarity of model-based identification strategies while leveraging the flexibility and generalization power of ML algorithms. Unlike conventional econometric models that rely heavily on parametric assumptions, or pure ML approaches that optimize prediction without inference guarantees, this hybrid literature aims to formalize data-adaptive structural estimators that are consistent, efficient, and interpretable for policy evaluation.

A review of 62 high-impact publications (2015–2024) from Web of Science reveals that this integration effort is centered on several converging themes. As shown in the VOSviewer map (Figure 5), the dominant topics include causal inference, propensity score matching, double robust estimation, instrumental variables, and generalized random forests, often co-appearing with applied domains such as healthcare, education, energy policy, and digital platforms. On the methodological side, big data analytics, deep learning, bias reduction, and heterogeneous treatment effects are frequently mentioned alongside foundational statistical terms like efficiency, identification, and confidence intervals. Meanwhile, the CiteSpace timeline (Figure 6) highlights three major phases in the field's intellectual trajectory: an early emphasis on efficient semiparametric estimation (2015–2018), a transition to policy-aware machine learning models (2019–2021), and a recent surge in causal machine learning for complex treatment regimes (2022–2024).

One influential line of research focuses on embedding graphical causal models or potential outcomes frameworks into ML

pipelines. For example, Louizos et al. (2017) introduced Causal Effect Variational Autoencoders, which use latent variable modeling to disentangle treatment effects from confounding. Shalit et al. (2017) proposed Counterfactual Regression Networks for learning individualized treatment effects under covariate shift. These architectures mark a significant shift away from black-box prediction toward model-aware architectures that respect treatment assignment mechanisms. Meanwhile, in the econometrics tradition, researchers such as Pearl (2009) and Bareinboim & Pearl (2016) advocate for structural causal diagrams (SCMs) that formalize identification via do-calculus, and provide blueprints for combining data with domain knowledge to estimate causal parameters under selection bias or mediation.

The synthesis of these approaches has led to powerful new estimators. Notably, generalized random forests (Athey et al., 2019) enable partition-based estimation of conditional average treatment effects (CATE), while causal boosting and meta-learners (e.g., T-learner, X-learner) adapt traditional ensemble learners for policy evaluation tasks. These innovations often outperform classical regression models in terms of bias-variance tradeoffs, especially when treatment effects are highly heterogeneous and complex interactions abound.

However, this emerging literature also reveals key tensions between structural rigor and algorithmic complexity. First, while ML tools excel at approximating nuisance functions, they often lack explicit representation of causal assumptions, making it difficult to evaluate identification strength or falsify key assumptions (e.g., unconfoundedness). Second, structural models offer transparency and counterfactual interpretability but are computationally fragile in high-dimensional settings, particularly under limited overlap or weak instruments. Finally, the lack of a unified inferential theory makes it difficult to combine frequentist inference with black-box learners, especially when interpreting uncertainty around causal estimates.

The proposed S-DIDML framework in this paper addresses these gaps by integrating Difference-in-Differences identification logic, double machine learning estimation, and domain-aware structural assumptions into a unified design. It maintains clear identification through parallel trends logic and temporal contrasts, enhances estimation robustness using orthogonalized residualization and cross-fitting, and enables high-dimensional adjustment through supervised ML. As such, S-DIDML contributes to the literature by offering a transparent, scalable, and statistically valid framework for policy evaluation with heterogeneous effects and complex treatments.

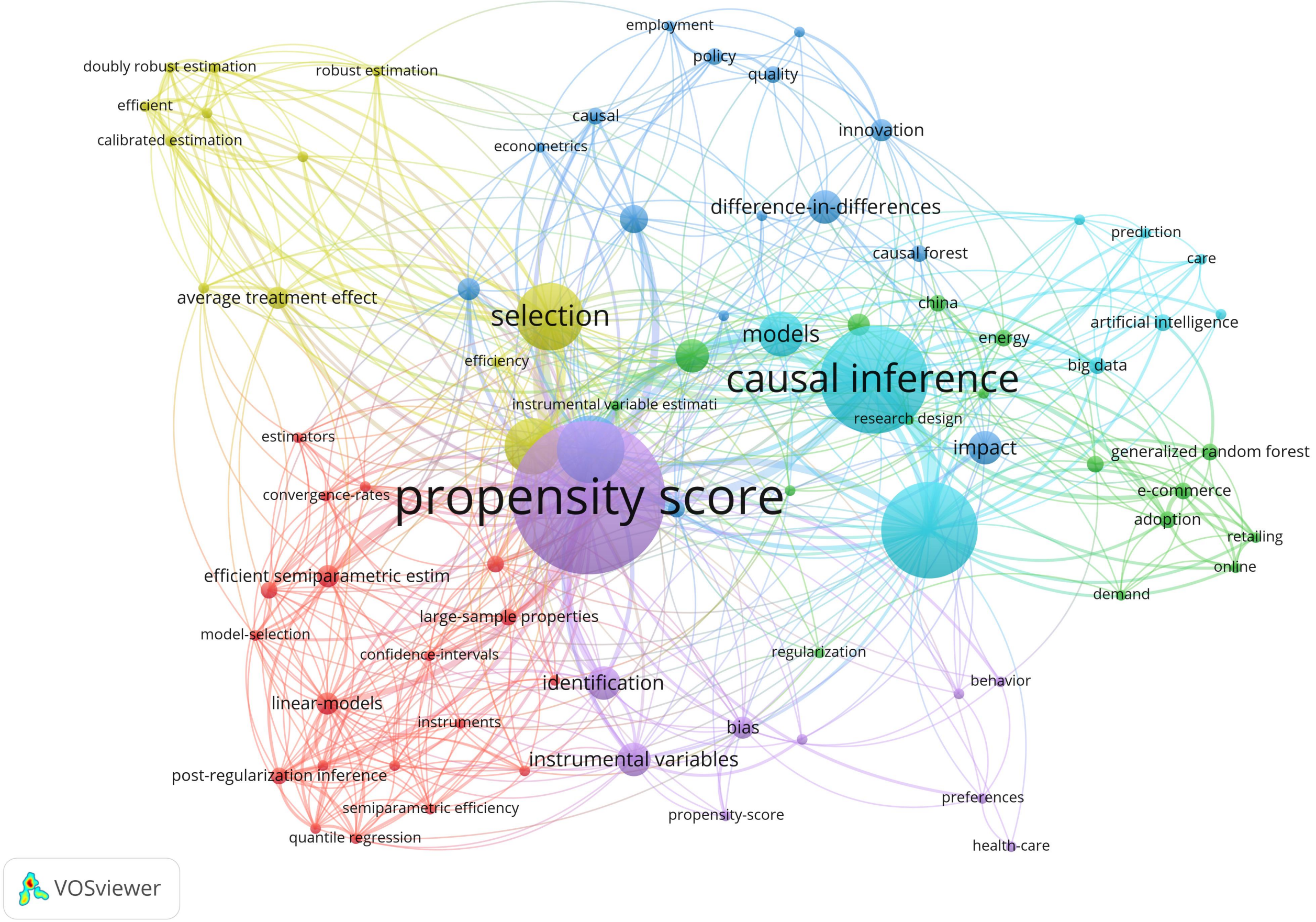

*Figure 5DDML & DID Co-occurrence Cluster Map of Key Research Terms*

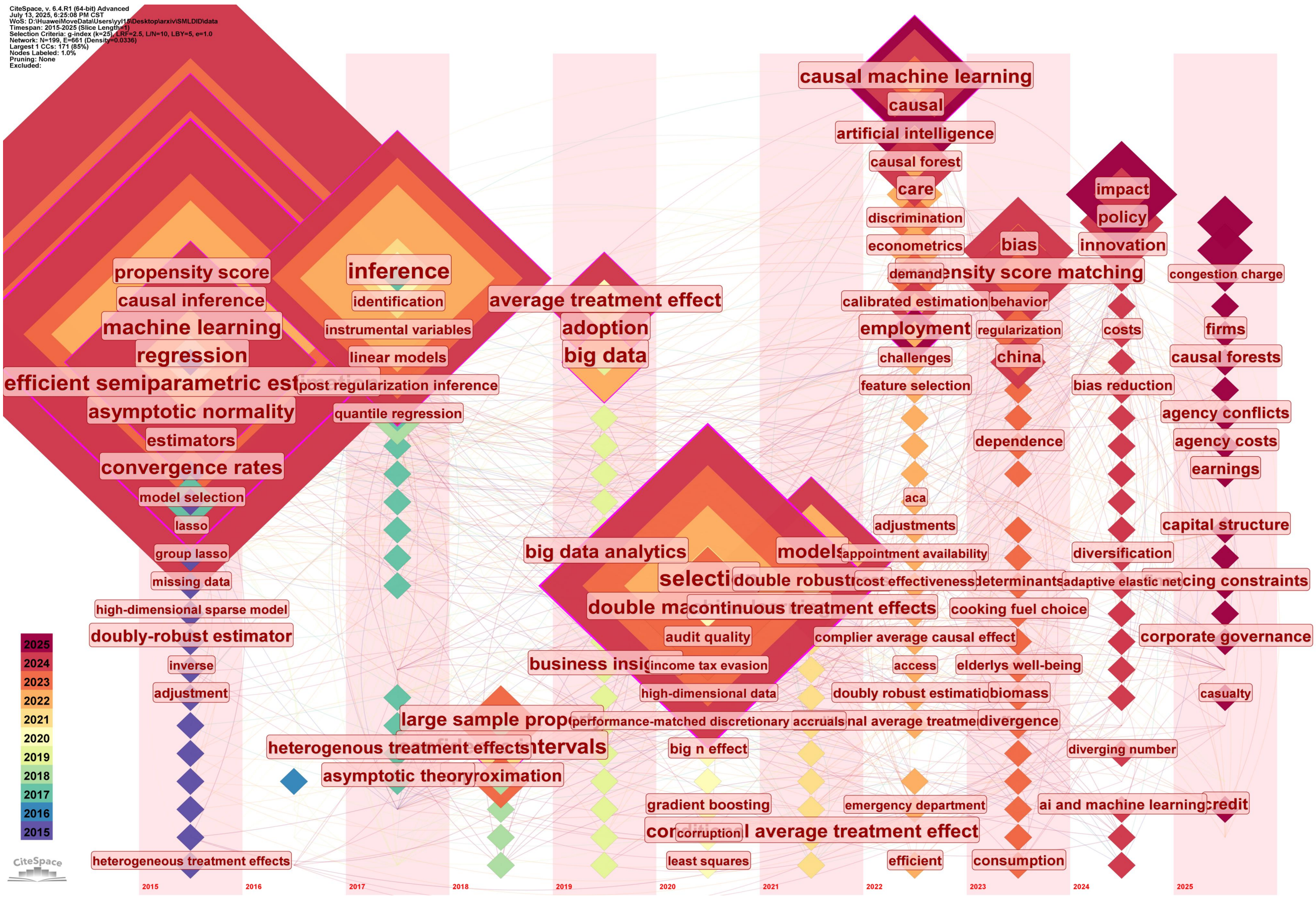

*Figure 6 Keyword Temporal Evolution Diagram of DDML&DID Integration Research*

## 3.Gaps and Research Needs

**A brief commentary is provided in the following table 1.**

*Table 1 Comparison Table of Characteristics of Different Methods*

| Method category | Core limitations | Performance | References |
|---|---|---|---|
| Machine learning(ML) | Lack of interpretability and structural modeling capability | Black-box models are difficult to interpret causal pathways; lack economic behavioral constraints or theoretical assumptions; and are non-analyzable for decision-making processes. | Breiman (2001); Chernozhukov et al. (2018); Molnar (2022) |
| Difference-in-Differences (DID) | Unable to handle high-dimensional covariates and complex heterogeneity structures | High-dimensional conditions lead to model instability; Assumptions (such as parallel trends) are difficult to verify; Unable to naturally identify multi-group, multi-time-point | Goodman-Bacon (2021); Roth et al. (2022) |
| | | heterogeneous effects | |
| Double/Debiased ML (DML/DDML) | Lack of structural embedding and multi-period adaptability | Assuming cross-sectional data is being processed, it is difficult to handle multi-period policies and temporal dynamics; divorced from the context of economic theory, the explanatory power is weak. | Chernozhukov et al. (2018); Imai & Kim (2021); Kennedy (2022) |
| HTE Estimation | Structural interpretation is lacking, and the mechanism transparency is low. | Causal Forests / Meta-Learners can detect heterogeneity but cannot explain the source of differences; they are prone to producing a "mechanism void" in policy interpretation bias. | Wager & Athey (2018); Künzel et al. (2019); Heckman & Vytlacil (2005) |

**3.1Machine Learning in Causal Inference: Challenges of Interpretability and Weak Structural Assumptions**

While machine learning (ML) techniques have demonstrated remarkable success in high-dimensional prediction tasks, their application in causal inference remains fraught with fundamental limitations—particularly regarding interpretability and structural grounding. This tension arises because most supervised ML algorithms are optimized for predictive accuracy rather than causal identifiability, leading to what Pearl (2009) famously described as "the algorithmization of association, not causation."

A key challenge is that ML models typically operate under agnostic data-generating assumptions, lacking a clear mapping to domain-relevant structural models. As Athey and Imbens (2019) emphasize, "machine learning excels at reducing prediction error but often leaves causal structure unspecified," which means treatment effects may be detected without being explainable in terms of mechanisms, counterfactual logic, or institutional design. For instance, while tree-based learners like random forests or gradient boosting can detect treatment heterogeneity, they do not clarify whether such heterogeneity is driven by observed policy variation, unmeasured confounding, or spurious interactions.

The lack of structural constraints in ML poses a second risk: it impedes credible identification of causal parameters. Unlike parametric econometric models that explicitly encode assumptions—e.g., exclusion restrictions, monotonicity, or sequential ignorability—ML models often estimate flexible functions without guiding restrictions. As noted by Louizos et al. (2017), this can lead to high-variance estimators or misleading results, particularly when treatment assignment is non-random or when confounding is only partially observed. Moreover, when ML is applied naïvely in causal contexts, it often violates the orthogonality conditions necessary for valid inference, unless careful debiasing or orthogonal score construction is used (Chernozhukov et al., 2018).

Interpretability is a closely related concern. As highlighted by Kitson (2025), although techniques such as SHAP or LIME provide local approximations for black-box models, they "do not constitute structural explanations of the data-generating process" and cannot substitute for a formal causal model. The absence of counterfactual semantics or policy-relevant parameters in standard ML frameworks makes it difficult to justify findings in regulatory, legal, or institutional decision-making.

Collectively, these limitations point to a broader epistemological issue: without structural assumptions, causal statements derived from ML risk being descriptive rather than explanatory. This weakens the scientific value of such analyses, particularly in social science domains that rely heavily on theoretical grounding, historical context, and institutional realism.

To address these concerns, emerging literature proposes combining ML with formal structural inference frameworks, such as Structural Equation Models (SEMs) (Pearl, 2009), structural score functions (Newey & Robins, 2018), or graphical causal modeling (Bareinboim & Pearl, 2016). However, these integrations remain nascent and have yet to be widely adopted in empirical work, particularly in policy evaluation with staggered or longitudinal designs.

### 3.2Limitations of Classical Difference-in-Differences: Covariate Dimensionality and Identification Fragility

The Difference-in-Differences (DID) framework has become a foundational method in applied econometrics for estimating causal effects from observational panel data. Its core appeal lies in its intuitive identification strategy, which relies on comparing outcome trends between treated and control groups, assuming the so-called parallel trends in the absence of treatment. However, despite its widespread use, recent theoretical and empirical developments have highlighted critical limitations of DID, particularly in the face of high-dimensional covariates, treatment timing heterogeneity, and violations of baseline assumptions.

One central issue concerns covariate control and model misspecification. Traditional DID models are typically estimated using two-way fixed effects (TWFE) regression, where control variables are either excluded or incorporated linearly. In practice, this limits the ability of DID to adjust for nonlinear, time-varying, or high-dimensional confounders—precisely the kinds of complexities prevalent in modern administrative, firm-level, or geospatial data. As Abadie (2005) observed, even moderate deviations from parametric assumptions in the outcome model can result in biased estimates. These risks are magnified when the covariate space expands or when interactions between covariates and treatment status are neglected.

Moreover, as Goodman-Bacon (2021) demonstrates, TWFE implementations in staggered adoption settings can lead to negative weighting of treatment effects, which introduces bias and undermines causal interpretability. This phenomenon occurs because the estimator aggregates over comparisons across different timing groups, some of which may act as implicit controls for others. In such cases, the estimate no longer represents a clear causal contrast between treated and untreated units, especially if treatment effects are dynamic or heterogeneous across cohorts. These insights have led to the development of more robust DID estimators, such as those proposed by Callaway and Sant'Anna (2021), which explicitly account for variation in treatment timing and allow for group-time-specific treatment effects.

A further challenge lies in the robustness to trend violations. Although DID is often justified by informal graphical checks or pre-trend tests, these approaches suffer from low power and subjective interpretation. Roth (2023) critiques the overreliance on pre-trends as a robustness diagnostic and proposes formal methods to account for uncertainty in parallel trend assumptions, showing that even mild violations can substantially alter inference. Moreover, in many policy environments—such as rolling interventions or gradually implemented regulations—the very notion of a single pre-treatment trend is implausible, requiring more flexible and data-driven trend modeling strategies.

Finally, DID's traditional framework is ill-equipped to accommodate high-dimensional settings, where the number of potential covariates far exceeds sample size. Under such conditions, linear fixed-effects models become unstable or inapplicable, and post-hoc covariate balancing or matching procedures may fail due to poor overlap or extreme weights. Attempts to augment DID with machine learning for pre-processing (e.g., via propensity score estimation) are often modular rather than integrative, failing to embed ML into the estimation stage or maintain orthogonality necessary for valid inference.

In sum, while DID remains a powerful identification tool, its classical implementations face substantial limitations in modern empirical settings characterized by high-dimensional data, staggered treatments, heterogeneous effects, and limited trend credibility. These limitations motivate the development of structurally grounded, machine-learning-enhanced DID frameworks—such as the S-DIDML estimator proposed in this study—that preserve the logic of temporal comparison while enhancing robustness, flexibility, and theoretical coherence.

### 3.3Double Machine Learning: Weak Integration with Economic Structure and Limited Temporal Flexibility

Double Machine Learning (DML) has emerged as a powerful framework for causal inference in high-dimensional settings. By leveraging orthogonal moment conditions, sample-splitting, and cross-fitting, DML enables consistent estimation of treatment effects even when nuisance functions (e.g., propensity scores or outcome regressions) are estimated using complex, flexible machine learning methods (Chernozhukov et al., 2018). However, despite its strong statistical foundation, DML remains structurally minimalistic and temporally constrained, making it ill-suited for many empirical contexts encountered in economic policy research.

One key limitation is DML's reliance on cross-sectional or static treatment settings. The majority of DML implementations are developed under the assumption of a single treatment decision per unit, which does not naturally extend to multi-period treatments, staggered adoption, or event-time heterogeneity—scenarios common in education, tax, environmental, or labor policy studies. As highlighted by Imai and Kim (2021), applying DML naively in panel data contexts can lead to biased estimates unless the temporal structure of treatment assignment and potential outcomes is properly accounted for. This restricts the applicability of standard DML in real-world settings where treatments unfold over time and responses are dynamic.

Furthermore, while DML provides flexibility in estimating nuisance functions, it does not by itself guarantee structural interpretability. Most DML applications are implemented without integrating formal economic assumptions—such as rational behavior, instrumental monotonicity, policy discontinuities, or selection mechanisms—into the estimation architecture. As a result, DML estimators may identify local causal effects without explaining their theoretical basis or economic implications.

Heckman and Pinto (2019) argue that such "structure-free" approaches risk producing effects that are "statistically significant but policy-irrelevant," particularly in domains requiring behavioral modeling or institutional contextualization.

In addition, DML's inference guarantees depend critically on certain orthogonality and rate conditions, which may break down in settings with low overlap, complex treatment interactions, or feedback loops over time. For example, Kennedy (2022) shows that conventional cross-fitting and orthogonalization procedures can perform poorly when the treatment assignment mechanism itself is endogenous to outcomes (e.g., performance-based subsidies or policy-induced behavioral changes). This further underscores the need for structural restrictions to ensure causal stability and policy extrapolation.

Another practical challenge lies in embedding DML within quasi-experimental designs, such as regression discontinuity, instrumental variables, or Difference-in-Differences frameworks. While recent work attempts to extend DML to instrumental variables (Chernozhukov et al., 2022) or event studies (Roth and Sant'Anna, 2023), these approaches are still in early stages and often lack the clarity and transparency associated with classical structural designs. The tension between statistical optimality and domain-specific identification logic remains unresolved.

Collectively, these limitations suggest that while DML is a powerful statistical tool, its effectiveness in economic and social science applications depends on its integration with structural frameworks that reflect timing, group heterogeneity, policy rules, and institutional mechanisms. Without such integration, DML risks becoming a technically sophisticated but substantively opaque estimation strategy.

### 3.4Structural Deficits in Heterogeneous Treatment Effect Analysis

The quest to uncover heterogeneity in treatment effects (HTEs) has grown rapidly across disciplines such as economics, political science, and public policy. However, the mainstream methodological paradigms—whether model-based subgroup analysis, decision-tree partitioning, or meta-learners—have been largely non-structural in nature, lacking integration with formal causal models that characterize behavior, mechanisms, or equilibrium responses. This structural deficit limits both the interpretability and transportability of HTE estimates across contexts.

Most machine learning approaches to heterogeneity estimation, such as Causal Forests (Wager & Athey, 2018) or meta-learners like T-learner or X-learner (Künzel et al., 2019), excel at data-adaptive partitioning but fail to encode economic or institutional structure. Their emphasis lies in function approximation rather than model-based identification, leading to what Khaled et al. (2025) describe as a "black-box heterogeneity explosion"—rich in surface-level variation, yet opaque in mechanism. Without incorporating constraints from economic theory (e.g., budget sets, behavioral responses, or equilibrium spillovers), these methods can generate misleading or policy-irrelevant inferences.

Moreover, the standard ML-based HTE frameworks are often agnostic to treatment effect pathways, such as mediation or interaction with latent policy regimes. As Banerjee & Veltri (2024) argue in the context of public policy evaluation, without integrating domain-specific structures or behavior-based causal graphs, "HTE estimates can be distorted by unmeasured confounding, model misspecification, or temporal misalignment." This is particularly relevant in longitudinal or staggered treatment settings where structural constraints like sequential ignorability, dynamic treatment effects, or latent policy phases need to be accounted for.

Structural heterogeneity modeling, by contrast, offers the promise of embedding causal variation within interpretable economic frameworks. For instance, instrumental variable heterogeneity (Carneiro et al., 2010), marginal treatment effect analysis (Heckman & Vytlacil, 2005), and structural latent-class models provide tools for interpreting how and why treatment effects differ across subpopulations. However, these approaches remain computationally intensive and less scalable to high-dimensional settings, leading to their underuse in empirical ML research.

A further structural limitation is the lack of policy extrapolation validity. ML-based HTE methods typically provide local estimates without guarantees of counterfactual stability under different policy environments. This undermines their use in ex-ante evaluations or general equilibrium simulations. As emphasized by Cobzaru (2025), "policy-relevant heterogeneity is not just a matter of detecting variation, but explaining it through causal structure that survives shifts in institutional or informational regimes."

Addressing these challenges requires a synthesis of structural causal modeling (SCM) and machine learning for heterogeneity discovery—a vision that remains largely unfulfilled. While recent work explores hybrid frameworks (e.g., Bayesian Additive Regression Trees with structural priors or interpretable surrogate models), these innovations are early-stage and lack standardized implementation for empirical social scientists.

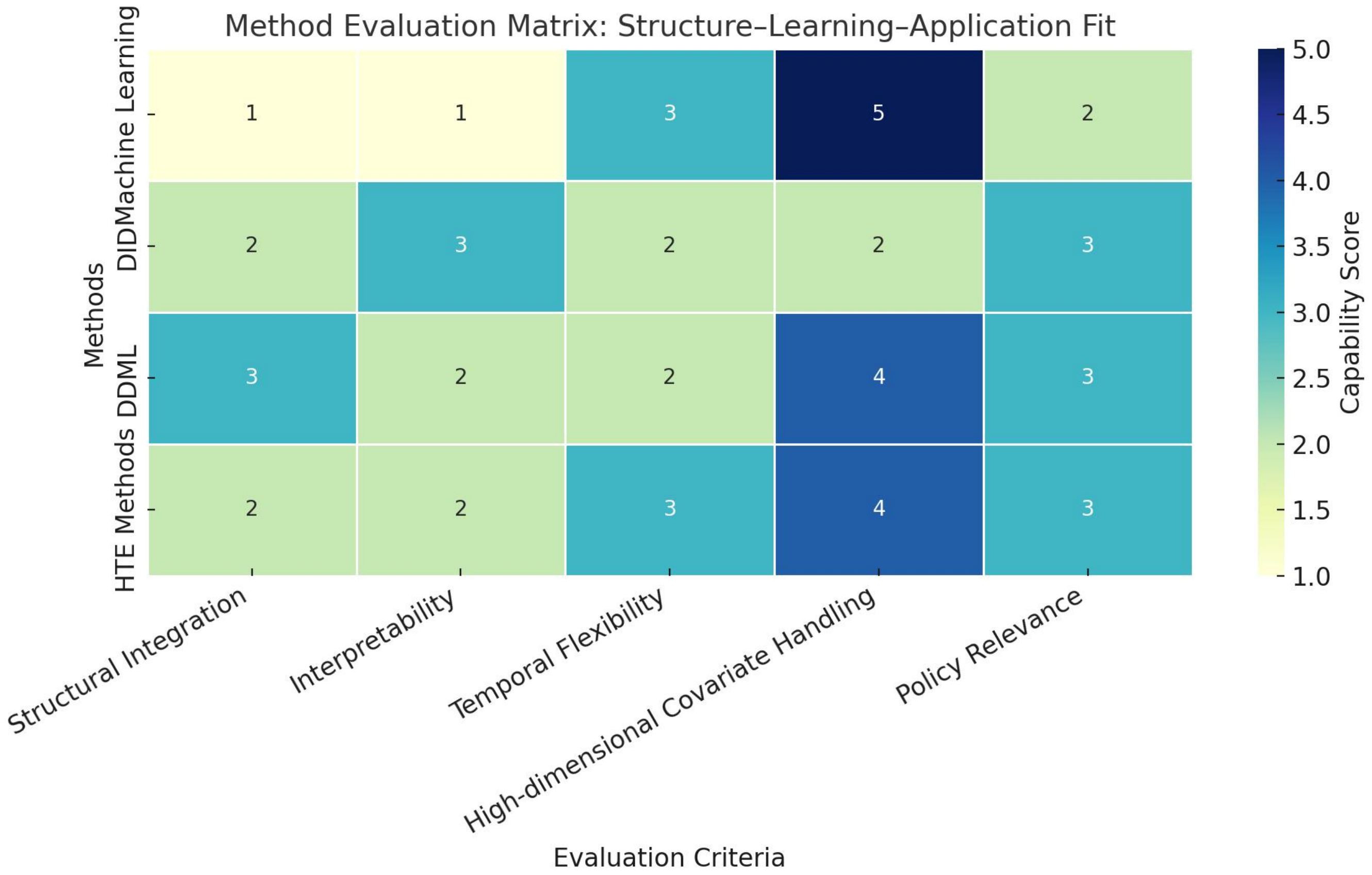

*Figure 7 Matrix heatmap estimation diagram of characteristics for various research methods*

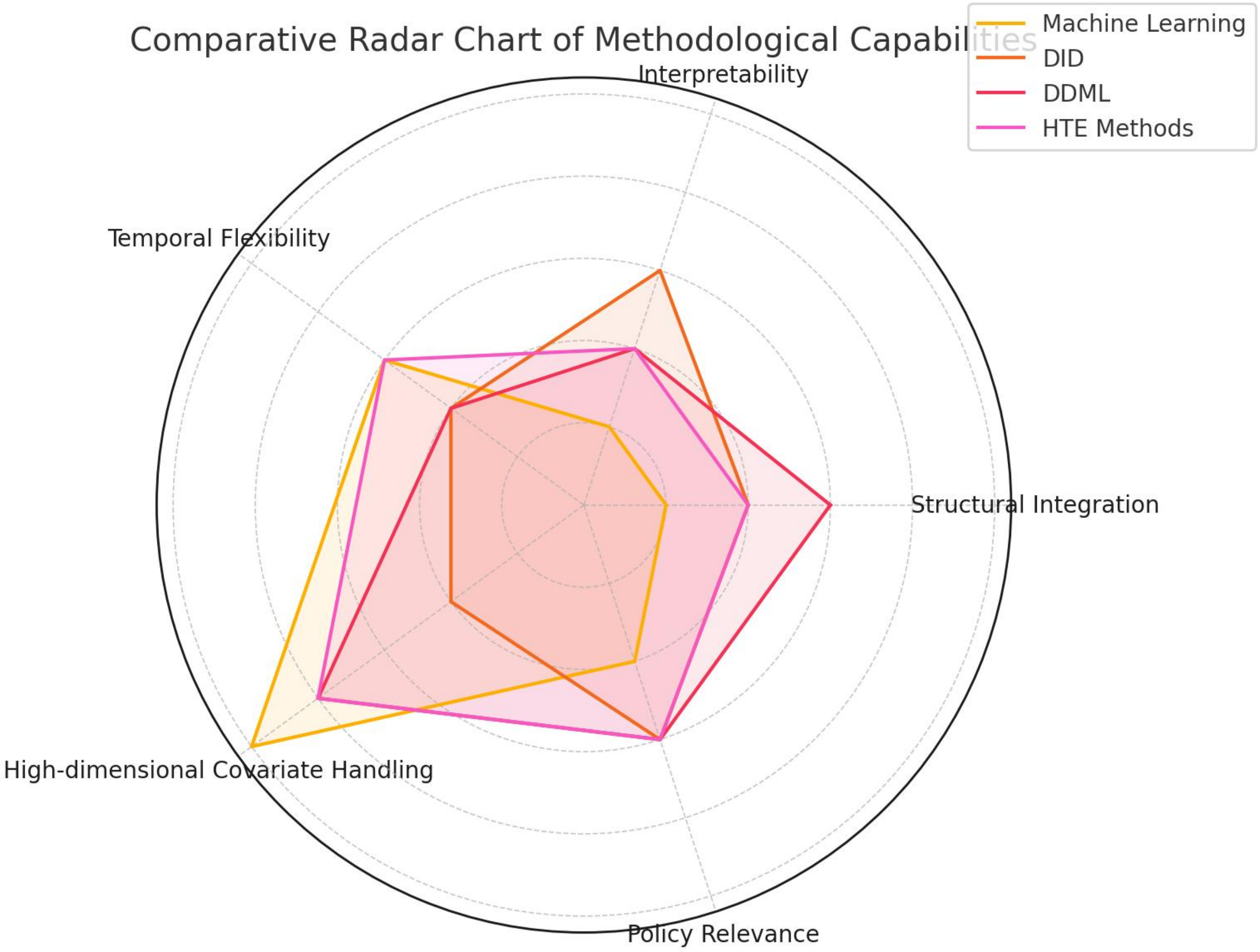

*Figure 8 Radar estimation diagram of the characteristics of various research methods*

## 4.The S-DIDML Framework: Structure, Principles, and Innovations

### 4.1Structure

The S-DIDML framework is constructed as a five-step semiparametric estimation pipeline, designed to combine the temporal identification logic of Difference-in-Differences (DID) with the residualization and orthogonalization principles of Double Machine Learning (DML). This design enables robust causal inference in high-dimensional, staggered-treatment settings while preserving interpretability grounded in structural counterfactual logic.

Step 1: Policy Exposure Encoding and Panel Structuring
The initial step involves transforming raw data into a well-defined panel format. Treatment exposure variables $D_it$ are encoded to reflect group-level policy adoption, including information on timing and staggered rollout. Cohort indicators $g(i)$and time indicators t are constructed, enabling the identification of treatment dynamics across subgroups and time periods. This step reproduces the structural basis of traditional DID under staggered adoption scenarios.

Step 2: High-dimensional Nuisance Estimation via Machine Learning
To control for confounding in high-dimensional settings, flexible machine learning models are used to predict the outcome and treatment assignment based on covariates. Specifically, two nuisance functions are estimated:

$$g(X_it)\approx E[Y_it|X_it],m(X_it)\approx E[D_it|X_it]$$

These models are fitted using cross-fitting to ensure orthogonality and to mitigate overfitting. A wide range of supervised ML

methods (e.g., random forests, boosting, neural networks) can be applied at this stage, depending on the structure of $X_i t$.

Step 3: Double Residualization of Outcome and Treatment Variables
Following estimation, the observed variables are residualized as follows:

$$\tilde{Y}_i t = Y_i t - \hat{g}(X_i t), \tilde{D}_i t = D_i t - \hat{m}(X_i t)$$

This double residualization process yields outcome and treatment variables that are orthogonal to the high-dimensional covariates $X_i t$, thereby satisfying the Neyman orthogonality condition necessary for valid second-stage inference.

Step 4: Structural DID Estimation on Residualized Quantities
The core causal effect is estimated by regressing the residualized outcome $\tilde{Y}_i t$ on the residualized treatment indicator $D_i t$, while incorporating group and time fixed effects. This regression preserves the cohort-time structure of DID and allows for dynamic, group-specific treatment effects:

$$\tilde{Y}_i t = \tau_g, t \cdot \tilde{D}_i t + \alpha_g + \lambda_t + \varepsilon_i t$$

Depending on the empirical setting, estimators such as Callaway–Sant'Anna (2021) or Sun & Abraham (2021) can be employed to estimate average or event-time treatment effects.

Step 5: Aggregation, Uncertainty Quantification, and Robustness Checks
Estimated group-time treatment effects $\tau_g$,t are aggregated into overall ATT or dynamic treatment effect curves. Standard errors are obtained using cross-fitting–compatible variance formulas or nonparametric bootstrap methods. Finally, robustness is assessed via falsification tests (e.g., placebo interventions), checking for pre-trend violations, and assessing overlap conditions.

**4.2Principles**

The S-DIDML framework is grounded in three foundational principles that jointly ensure the validity, robustness, and generalizability of causal estimates in complex, high-dimensional, and staggered policy evaluation settings. These are: (1) structural identification, (2) double orthogonalization, and (3) semiparametric flexibility. Together, these principles position S-DIDML as a bridge between the interpretability of quasi-experimental methods and the statistical power of modern machine learning.

(1) Structural Identification through Cohort–Time Designs
At its core, S-DIDML preserves and enhances the temporal and group-based identification logic of the Difference-in-Differences (DID) framework. Rather than assuming a homogeneous average treatment effect across time and groups, the method leverages cohort-specific treatment timing and calendar/event time decompositions to uncover heterogeneous dynamics. This aligns with recent advances in causal panel data literature (e.g., Callaway & Sant'Anna, 2021; Sun & Abraham, 2021) that advocate for group-time-specific treatment effect estimation to avoid invalid weighting and bias from negative weights.
By anchoring identification in group-time structure, S-DIDML ensures that treatment effect estimates retain clear causal meaning under parallel trends and overlap assumptions, even in the presence of high-dimensional controls.

(2) Neyman Orthogonality via Double Residualization
A central innovation in S-DIDML is its application of Neyman orthogonalization through double residualization. Both the outcome and treatment variables are residualized on high-dimensional covariates via flexible machine learning models. This design achieves two goals:
- It neutralizes first-stage estimation error from the ML nuisance models.
- It ensures that second-stage estimation (of τ) is robust to overfitting and consistent even when nuisance models are biased, as long as the orthogonality condition holds.
This principle is inherited from the Double Machine Learning (DML) literature (Chernozhukov et al., 2018), but S-DIDML

extends it by embedding it within a structural DID contrast, yielding interpretable and robust effect estimates.

(3) Semiparametric Flexibility and High-dimensional Adaptability
S-DIDML embraces a semiparametric structure, combining nonparametric modeling of nuisance components with parametric identification of treatment effects. This yields multiple advantages:
- It allows for flexible modeling of nonlinear relationships, complex interactions, and high-cardinality categorical variables.
- It preserves interpretability, as treatment effects are still defined structurally through the contrast between residualized treated and untreated units across groups and time.
- It enhances scalability, as the modular two-stage structure separates learning and estimation.
Such flexibility makes the estimator particularly well-suited for modern datasets in economics, policy science, and social research, where covariate dimensions are high, treatment timing is staggered, and classical parametric models may be misspecified.

**4.3Innovations**

The S-DIDML framework introduces a set of integrated innovations that extend the methodological frontier of causal inference at the intersection of economics, machine learning, and structural modeling. These innovations not only respond to the identified limitations of DID, ML, and DDML methods, but also actively synthesize their strengths into a coherent, interpretable, and generalizable estimation strategy. Three primary innovations distinguish S-DIDML:

(1) A Structural Reframing of DML within a Panel Counterfactual Logic
Unlike conventional DML frameworks, which are primarily cross-sectional and agnostic to treatment timing, S-DIDML embeds DML's orthogonalization and high-dimensional flexibility into a panel-based counterfactual framework rooted in staggered Difference-in-Differences designs. This reframing allows the estimator to maintain policy-relevant interpretability (e.g., ATT for dynamic cohorts) while absorbing the flexibility of machine learning.
This structural embedding is particularly critical for policy analysis, where treatment effects are expected to evolve across cohorts, time, and institutional settings.

(2) The Double Residualized DID Estimator (DR-DID): A Hybrid Estimation Strategy
S-DIDML introduces the Double Residualized DID (DR-DID) estimator, which generalizes the idea of regression adjustment and inverse-propensity weighting into a two-stage semiparametric framework. Unlike recent double-robust estimators (e.g., Sant'Anna & Zhao, 2020), DR-DID allows both the outcome and treatment models to be learned flexibly using machine learning algorithms, while retaining interpretable group-time contrasts from the DID literature.
Furthermore, the method employs cross-fitting to mitigate overfitting and guarantee Neyman orthogonality, thereby ensuring robustness of inference even under model misspecification or high-dimensionality.

(3) A Modular Pipeline Enabling Scalability, Adaptability, and Robustness
S-DIDML is implemented as a modular pipeline, where each step (panel structuring, ML nuisance estimation, residualization, structural regression, aggregation) is fully separable and extensible. This allows for:
- Scalability: Methods scale easily with increasing covariate dimensionality or large samples.
- Adaptability: Researchers can plug in modern ML methods (e.g., boosting, forests, transformers) in the first stage.
- Robustness: Built-in diagnostics (e.g., residual-based pre-trend tests, subgroup ATT tracking) improve reliability in empirical work.
This modular architecture makes S-DIDML a general-purpose framework, suitable for both confirmatory and exploratory policy evaluations in heterogeneous environments.

**5.Demonstrative Literature Applications: Where S-DIDML Can Be Used**

The S-DIDML framework is not only theoretically robust but also practically versatile. Its design enables immediate adoption in several streams of empirical literature, especially where conventional DID or DDML frameworks face limitations due to high-dimensional covariates, staggered policy timing, or heterogeneity in treatment effects. Below, we outline four thematic domains where S-DIDML can provide substantial improvements in causal identification and inference quality.

### 5.1 Labor Economics: Evaluating Active Labor Market Policies (ALMPs)

Many ALMP evaluations rely on DID or event-study approaches (e.g., Kluve, 2010; Card et al., 2018), often using limited covariates due to multicollinearity concerns. However, modern administrative labor datasets now include thousands of features (firm size, tenure, dynamic local shocks). S-DIDML allows for robust estimation of heterogeneous effects of job subsidies or training programs across firms, sectors, or worker types, while maintaining structural interpretability of ATT.

### 5.2 Education Policy: School Reform, Curriculum Changes, and Tracking

Educational reforms (e.g., extending school years, STEM incentives, curriculum realignments) are often evaluated via DID with state or district fixed effects. However, treatment rollout is usually staggered, and student-level data are high-dimensional. By orthogonalizing outcomes with rich baseline test scores, socio-demographics, and parental inputs, S-DIDML enables dynamic treatment effect estimation at both cohort and demographic subgroup levels.

### 5.3 Fiscal and Tax Policy: Estimating Behavioral Responses with Administrative Tax Data

Tax policy changes (e.g., earned income tax credits, marginal rate changes) exhibit rich staggered designs but require flexible modeling of income dynamics, deductions, and family structure. Traditional regression-based DID models are poorly suited for such settings. S-DIDML accommodates high-dimensional pre-tax characteristics and allows for precise subgroup inference on labor supply elasticities or compliance behavior.

### 5.4 Environmental and Urban Policy: Evaluating Green Subsidies and Urban Interventions

Many environmental interventions (e.g., subsidies for electric vehicles, zoning regulations, pollution controls) are staggered and differ in intensity across space and time. Treatment heterogeneity is fundamental, and the policy environment is often rich in covariates (weather, geography, firm-level pollution histories). S-DIDML can flexibly estimate subgroup effects (e.g., by income decile or industry), while adjusting for spatial autocorrelation and nonlinearity.

### 5.5 Opportunities in Heterogeneity Mapping and Welfare Simulation

Beyond direct ATT estimation, S-DIDML can be embedded into policy simulation pipelines, enabling credible counterfactual mapping of treatment gains across the covariate space. For instance, it can assist in identifying which demographic groups benefit most from job guarantees or minimum wage hikes, using machine learning for heterogeneity partitioning while ensuring DID identification integrity.

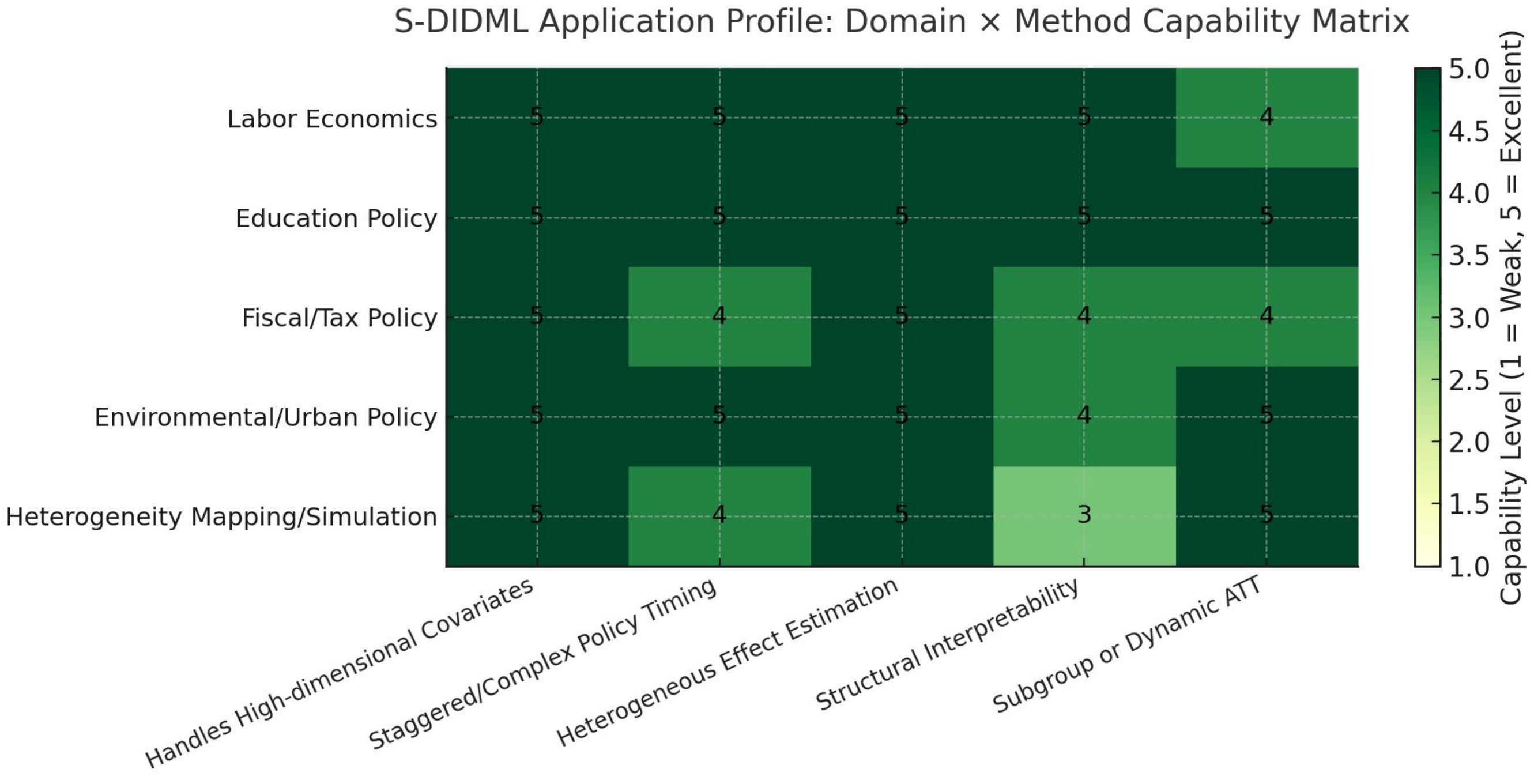

*Figure 9 Matrix heatmap estimation diagram of S-DIDML framework applications across different domains*

## 6.Limitations and Future Methodological Directions

While the S-DIDML framework provides a structurally grounded and high-dimensionally adaptable solution to causal inference in staggered treatment settings, several limitations and open challenges remain. Recognizing these boundaries is critical both for theoretical transparency and for guiding future innovations in applied econometrics.

### 6.1 Residualization Assumes Sufficient Overlap and ML Consistency

S-DIDML relies on accurate first-stage estimation of nuisance functions $\hat{g}(X_i t)$ and $\hat{m}(X_i t)$, which are used to construct orthogonal residuals. However, this process is only valid under conditions of sufficient covariate overlap (i.e., common support) and the consistency of ML estimators. In sparse regions of the covariate space or in samples with limited diversity, residualization may introduce instability, leading to inflated variance or attenuation bias.

Future work may consider integrating covariate balancing methods or adaptive sample trimming techniques to improve robustness in low-overlap regions.

### 6.2 Imperfect Handling of Spillovers and Interference

As with all unit-level causal designs, S-DIDML assumes no interference between units (SUTVA). However, in real-world policy settings—particularly in urban, environmental, or networked domains—spillover effects are common. The current framework does not model such interdependencies explicitly.

A fruitful extension would incorporate spatial or network-based residualization layers, or develop group-level S-DIDML variants that incorporate correlated treatment effects across units.

### 6.3 Limited Theoretical Exploration of Longitudinal Treatment Heterogeneity

While the framework accommodates staggered adoption and dynamic ATT, it currently assumes relatively stable causal structures across time. In contexts where treatment effects evolve nonlinearly (e.g., due to adaptation, fatigue, or saturation), a single-layer residualization may be insufficient.

Further methodological research could develop multi-period orthogonalization strategies, or state-dependent treatment dynamics models, integrated with dynamic panel machine learning.

### 6.4 Absence of Integrated Inference for Simultaneous Heterogeneity Testing

S-DIDML provides estimates of subgroup-specific ATT, but lacks a unified inference procedure to test for systematic heterogeneity across subpopulations. Moreover, there is no built-in method for controlling false discovery rates (FDR) or multiplicity when comparing many groups or time periods.

Methodological advances are needed in high-dimensional post-DID inference, combining ideas from selective inference, multiple testing theory, and partial identification.

### 6.5 Software and Computational Challenges in Large-scale Implementation

While conceptually modular, S-DIDML involves multiple cross-fitting and ML procedures that may be computationally expensive. Reproducibility and transparency require well-structured software pipelines and diagnostic tools.

Developing open-source toolkits (e.g., in Python, R, or Julia) with user-friendly API design and pre-built diagnostics (e.g., balance checks, placebo tests) is an important priority for dissemination.

In sum, S-DIDML lays a foundational bridge between structural causal identification and machine learning. Its limitations do not reflect methodological weaknesses, but rather opportunities for future refinement. As data structures and computational environments continue to evolve, we foresee a vibrant space for expanding this framework into nonparametric structural models, spatially-aware causal designs, and interpretable ML pipelines tailored to policy inference.

## 7.Conclusion

This paper introduces the S-DIDML framework—a structural, semiparametric estimator designed to bridge the interpretability of Difference-in-Differences (DID) methods with the high-dimensional flexibility of Double Machine Learning (DML). In doing so, we aim to provide applied researchers in economics and the social sciences with a unified, theoretically grounded, and computationally feasible approach to estimating heterogeneous treatment effects in staggered policy contexts.

We began by identifying a set of unresolved challenges in modern causal inference: the limited interpretability of pure machine learning estimators, the instability of conventional DID methods under high-dimensional controls, and the restricted scope of existing DDML estimators in handling multiple treatment periods or complex policy rollout. Through an extensive literature review across DID, DDML, and structural ML, we demonstrated the methodological need for an integrated solution.

To address these limitations, S-DIDML proposes a five-step estimation pipeline: (1) panel structuring and treatment timing encoding, (2) flexible nuisance estimation using ML, (3) double residualization for Neyman orthogonality, (4) structural DID regression for interpretable group-time effects, and (5) aggregation and robustness analysis. Each step is modular, theoretically motivated, and designed for transparency and scalability.

We articulated the framework's principles—structural identification, orthogonality, and semiparametric adaptability—and illustrated its potential across key empirical domains including labor policy, education, taxation, and environmental regulation. Furthermore, we engaged critically with its current limitations, such as reliance on overlap, lack of interference modeling, and the need for unified subgroup inference. These issues mark important frontiers for future research in causal machine learning.

S-DIDML is not intended to replace either traditional quasi-experimental methods or deep learning-based prediction tools. Instead, it acts as a conceptual and computational bridge: retaining causal interpretability grounded in economic theory, while leveraging the modeling capacity of modern ML for complex data environments. As empirical researchers face ever-larger datasets and increasingly heterogeneous policy designs, such hybrid frameworks are crucial for producing credible, robust, and policy-relevant insights.

We envision S-DIDML not as a fixed model but as a flexible blueprint—one that invites further theoretical refinement, software development, and empirical adaptation. Its goal is not merely to improve estimation, but to foster a new generation of structurally informed, statistically rigorous causal inference in the high-dimensional era.

## 8.Fund

This research is supported by the project 'Path of Government Financial Input Restructuring to Promote High Quality Development of Education'（Project No.2025279）of the Canal Cup Extracurricular Academic Science and Technology Fund for College Students of Zhejiang University of Technology.

## 9.Conflict of interest

The authors declare that there are no conflicts of interest regarding the publication of this article.

## 10.References